# Quantum-AI empowered Intelligent Surveillance: Advancing Public Safety Through Innovative Contraband Detection

**Syed Atif Ali Shah**[1,2] *****, welcomeatif@yahoo.com, Orcid ID: https://orcid.org/0000-0002-4275-9731
**Nasir Algeelani** [2], nasir.ahmed@mediu.edu.my
**Najeeb Al-Sammarraie** [2.], dr.najeeb@mediu.edu.my
[1] Department of CS, Al-Madinah International University,
[2] Department of CS, Al-Madinah International University, Malaysia.

**Abstract**
Surveillance systems have emerged as crucial elements in upholding peace and security in the modern world. Their ubiquity aids in monitoring suspicious activities effectively. However, in densely populated environments, continuous active monitoring becomes impractical, necessitating the development of intelligent surveillance systems. AI integration in the surveillance domain was a big revolution, however, speed issues have prevented its widespread implementation in the field. It has been observed that quantum artificial intelligence has led to a great breakthrough. Quantum artificial intelligence-based surveillance systems have shown to be more accurate as well as capable of performing well in real-time scenarios, which had never been seen before. In this research, a RentinaNet model is integrated with Quantum CNN and termed as Quantum-RetinaNet. By harnessing the Quantum capabilities of QCNN, Quantum-RetinaNet strikes a balance between accuracy and speed. This innovative integration positions it as a game-changer, addressing the challenges of active monitoring in densely populated scenarios. As demand for efficient surveillance solutions continues to grow, Quantum-RetinaNet offers a compelling alternative to existing CNN models, upholding accuracy standards without sacrificing real-time performance. The unique attributes of Quantum-RetinaNet have far-reaching implications for the future of intelligent surveillance. With its enhanced processing speed, it is poised to revolutionize the field, catering to the pressing need for rapid yet precise monitoring. As Quantum-RetinaNet becomes the new standard, it ensures public safety and security while pushing the boundaries of AI in surveillance.
**Keywords:** Quantum AI; Deep Learning; Quantum Deep Learning; CNN; QCNN; Intelligent Surveillance; Weapon detection.

## Introduction

Intelligent surveillance research has significant implications across various domains, including public safety, security, and law enforcement efforts. It can be employed in public spaces, transportation hubs, schools, and other crowded areas to identify and prevent potential threats, mitigating risks associated with armed violence and terrorist activities. By integrating such systems into law enforcement operations, authorities can identify and apprehend individuals carrying illegal contraband or harmful objects, reducing the occurrence of violent crimes. Mass shooting prevention can be achieved by detecting contraband in locations like schools and public events, while border security is essential for securing borders and preventing illegal contraband trafficking. Military settings can benefit from systems in identifying and neutralizing enemy threats, providing added protection for

---

troops. Airport and aviation security can be enhanced by detecting concealed contraband in carry-on luggage and other areas. Prisons and correctional facilities can also benefit from technologies, enhancing overall security for staff and inmates. AI-based image identification, infrared imaging, millimeter-wave scanners, and improved sensor systems are examples of technological advances in contraband detection.

Quantum computing is a field of computer science and physics that explores the principles and technologies underlying quantum mechanics to develop new types of computers. Traditional computers use bits that represent either a 0 or 1 to process and store information, while quantum computers use quantum bits or qubits, which can exist in multiple states simultaneously. This enables quantum computers to perform certain computations much faster than classical computers, especially for complex problems such as factorization and optimization. Based on ideas like quantum superposition, entanglement, and interference, quantum computing has the potential to transform several industries, including banking, drug research, encryption, and AI. However, there are still problems to be solved, such as creating dependable quantum hardware and creating new algorithms that make use of quantum computing. An emerging technology called quantum deep learning uses deep learning methods to solve complex issues in fields including image recognition, natural language processing, and drug development. To carry out deep learning tasks, this entails creating quantum algorithms and quantum computing hardware.

Quantum gates are used by the quantum circuits to process and alter the quantum states created by the encoding of data into them for deep learning tasks. Despite these difficulties, there is significant interest in quantum deep learning, and organizations and academics all around the world are looking at how it may lead to new developments in AI. In a hybrid CNN-QCNN, QCNN is used for later layers, such as classification or detection, while traditional CNN is used for earlier levels, like feature extraction and reduction. While quantum-inspired optimization algorithms like the Variational Quantum Eigensolver (VQE) or Quantum Approximate Optimization Algorithm (QAOA) can be used, this method can also be learned using conventional deep learning approaches like backpropagation. Quantum deep learning is a potential area of technology for the future since it can improve performance on some sorts of issues when the classical CNN and quantum methods are combined. For instance, the data may be preprocessed using the traditional CNN to extract relevant features, which can then be fed into the QCNN for additional processing and analysis.

In this research ethical problems are also addressed, with studies focusing on privacy concerns and potential biases in the deployment of Intelligent surveillance technology. Real-time warning and integration with surveillance systems provide a speedy reaction to possible threats. Overall research has far-reaching consequences for the safety and security of individuals and communities, but striking a balance between security measures and individual rights and privacy needs careful consideration of ethical and legal factors. Collaborations between researchers, law enforcement agencies, and policymakers are critical for the development and application of responsible Intelligent surveillance technologies.

## Literature review

Countries with large stocks of contraband (such as guns) have very high crime rates. According to several sources, the illegal acts have caused a variety of consequences, including murder, theft, destruction of infrastructure, and loss of billions of dollars(1).

Traditional CCTV cameras are used to monitor specific areas, but the surveillance technique is more manual (2). This scenario has been modified by deep learning. In this regard, researchers have developed many models for identifying contrabands(3). This article explores the various possibilities in detail. The research ranges from early contraband detection systems to the latest designs available(4). This journey began with manual methods and ended with fully automated intelligent systems(5). We started with a manual that incorporated weighted Gaussian mixtures(6), polarization signal-based methodologies, multiresolution mosaicism, three-dimensional (3D) computed tomography (CT), and the Haar cascade (7)method into our work. Machine learning approaches such as the Visual Background Extractor Algorithm, 3-layer ANNs used in conjunction with active mmWave radar (8), X-ray-based methodologies(9), and ANN-based MPEG-7 classifiers are of interest. With the introduction of SURF (Fast Robust features), Harris Interest Point Detector (HIPD)(10), and Fast Retina Keypoint, the era of Convolutional Neural Networks (FREAK) has begun (11). The arrival of various models was also announced to improve efficiency. Related to this, there are also many different scientific publications using deep convolutional networks and transfer learning technology(12). For example, his X-rays are used for classification and infrared for concealed contrabands(13). Recently, a special CNN model was announced that is not only accurate but significantly speeds up the detection of contraband in streaming video. These models include R-CNN (14), Fast R-CNN(15), Faster R-CNN(16), Inception, YOLO, VGG-Net, ZF-Net, and YOLO-V3. In addition to speed and accuracy, another important factor is complexity(17). This allows the model to run more smoothly on small devices such as smartphones, and he can use it in IoT apps to study such areas. RPN (Regional Proposal Network)(18), GoogleNet, SqueezeNet, HyperNet, RetinaNet, LeNet, AlexNet, ZFNet, GoogleNet, VGGNet, ResNet, Startup Model, ResNeXt, SENet, MobileNet V1/V2, X, NASNet, PNASNet, ENASNet, EfficientPointNet, MobiiliNet2, Inception ResNetV2, ResNet50 and other models have been developed. This study also includes the meaning of many terms used in the literature. B. "Complex Backbone" vs. "Light Bone", "Two-Phase Indicator" vs. "One-Phase Indicator", and Pros and Cons of Different Models. We note that our essays not only present research models (how they have changed over time) but also help researchers establish a solid foundation from which to start their research(19). This review discusses the evolution of Intelligent surveillance models and their performance in detecting guns and pistols(20). Traditional methods for firearm detection, such as X-ray technology (21) or millimetric wave imaging(22)(23), are expensive and impractical due to their reaction to all metallic items, including contrabands (24). However, deep learning has proven to be the most effective method of learning, with convolutional neural networks (CNNs) outperforming traditional methods(25). Transfer learning, which re-utilizes information from one domain to another related domain, is becoming popular. Other popular techniques include Scale-Invariant Feature Transform (SIFT), Rotation-Invariant Feature Transform (RIFT) (26), and Fast Retina Keypoint (FREAK)(23). Some initial work has been done to detect pistols in images, but it is not able to predict multiple pistols in one image (27). The Bag of Words Surveillance System (BoWSS) (28)algorithm was used to detect guns in images, and Faster R-CNN deep learning was used to detect a hand-held gun(29). However, the model can only detect and locate pistols, and it often fails to detect other types of contraband, such as machine guns(30). Fernandez et al. presented a new CNN model for detecting guns and knives from video surveillance systems, comparing it to GoogleNet and SqueezeNet(31). SqueezeNet had better performance in gun detection, while GoogleNet had better performance with knife detection (32). A sequential deep neural network-based approach was developed to resolve three major problems with automated BCG detection and learning(33). Real-time detection of pistols remains a challenge due to factors such as distance, visibility, type of pistol, scale, rotation, and shape(34). Advancements in these domains are being made to improve accuracy and performance in contraband detection(35).

## Dataset

For this research we have designed our dataset, to include the contrabander especially used in $3^{rd}$ world countries. Other datasets contain multiple types of armory but usually contain contrabands used in Hollywood movies or/and used in developed countries. When the same model is implemented in $3^{rd}$ world counties then it shows lower accuracy. To make our work acceptable worldwide, we developed our dataset that contains an armory used in all parts of the world. This research is mainly concerned with the types of armory used in street crimes like Shortrange Rifle, Shotgun, Pistols, and knives thus named it Street Crimes Arms dataset (SCAD). Each category contains almost 5000 images, hence contains 20000 (app) images in total. For smart training of the models, a dataset has gone through different alterations which include augmentation and normalization.

## Methodology

This research combines the power of Quantum computing with conventional Deep Learning (Convolutional Neural Network). Both technologies have their own strengths and weaknesses. Quantum computing is famous for tremendous speed but due to unavailability of hardware, usually not found in implementations. On the other hand though applications of Deep learning are widely used but lacking the speed in many real-time scenarios. Combination of both technologies reveal a new dimension, and thus suppresses their weaknesses. Nowhere, QCNN is combined with RetinaNet. Feature extraction part of RetinaNet is replaced by QCNN. Thus achieves the speed of Quantum Computing and accuracy of RetinaNet.

**RetinaNet**

RetinaNet is a state-of-the-art object detection algorithm that was introduced by researchers at Facebook AI Research in 2017. It is designed to solve the problem of detecting objects in an image, where the location and type of objects can vary widely. The key innovation of RetinaNet is the use of a novel loss function called Focal Loss, which addresses the issue of class imbalance in object detection. In object detection, there are typically many more negative examples (background) than positive examples (objects of interest). RetinaNet utilizes Focal Loss to reduce bias toward negative examples and focus on hard examples in the positive class. This helps the model to identify rare and important positive cases. To recognize objects of varying sizes and resolutions, RetinaNet's feature pyramid network (FPN) design mixes low-resolution and high-resolution characteristics. The model has gained cutting-edge performance on benchmarks like COCO and PASCAL VOC, making it a popular choice for object identification tasks in industrial and commercial applications.

**Quantum-RetinaNet**

The core functionality of the RetinaNet is divided into two main parts. The first one called feature extractor, which deals with the extraction of features. The second one is called task-specific networks, which is responsible for Classification and Bounding Box. The first part i.e the feature extractor; that uses convolution and pooling layers to extract features which is a time consuming process. In Quantum-RetinaNet this time is reduced using Quantum-Convolutional Neural Network.

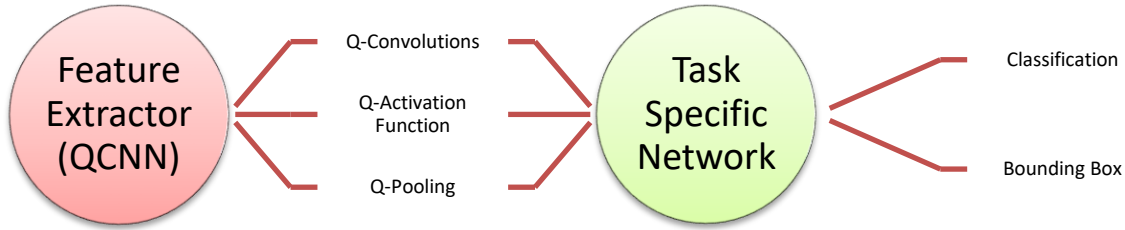

**Figure 1: The basic architecture of the RetinaNet, first part deals with the extraction of features and second one is task-specific networks.**

Fig 1 describes the basic working of the Q-RetinaNet. The Q-RetinaNet model converts an image into a vector of corresponding bits for quantum computing. Qubits are fabricated, and feature extraction is performed using QCNN. These features are then forwarded to Task Specific Networks, which draw bounding boxes and detect desired objects.

*Feature Extraction using Qonvolutional Neural Network*

A Quantum Convolutional Neural Network or Qonvolutional Neural Network uses quantum circuits to conduct convolutional operations on input data . Using methods like amplitude encoding or qubit encoding, the procedure entails converting the incoming data into a quantum state. To identify certain characteristics, such as edges or textures, convolutional filters are used. After that, the data is pooled using a pooling layer, which lowers the dimensionality while maintaining crucial properties, as shown in the fig 2.

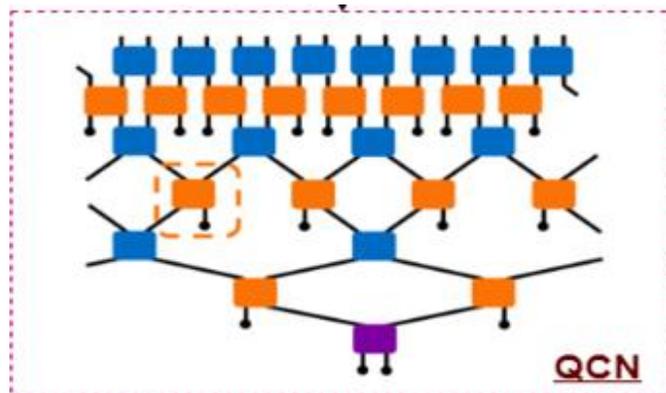

**Figure 2: QCNN uses quantum circuits to conduct convolutional operations on input data, using amplitude encoding or qubit encoding to convert incoming data into a quantum state**

For classification tasks, the output is processed utilizing additional layers of quantum circuits, employing methods such as variational quantum classifiers or quantum SVMs. To improve network performance, the parameters of the quantum circuits are improved using strategies like quantum gradient descent or variational approaches.

*Bits to Qubits*

The data contained in a bit is encoded into the single qubit's quantum state by transforming it from a classical bit to a quantum bit. A qubit is a type of quantum bit that may exist in several states at once as opposed to a classical bit, which can only exist in one state at a time. Initialize the qubit in the |0 state, which corresponds to a classical bit's 0 state, to encode a classical bit into it. Apply the X gate to change the qubit from its |0| state to its |1| state if the classical bit is 1. The qubit's final state corresponds to the conventional bit's encoded form. Depending on the exact application, there are several additional techniques to convert conventional bits into qubits.

*Quantum Convolutions*

Quantum convolutions are implemented using quantum circuits, gates, and quantum algorithms. While gates, such as the Hadamard-Walsh transform gate for discrete cosine transform (DCT) and discrete wavelet transform (DWT) operations, apply the convolutional filter directly to a quantum state, circuits employ quantum Fourier transforms (QFTs) to change a quantum state. Quantum algorithms are created to work on quantum data and carry out various tasks, including quantum convolution, such as the quantum singular value transformation (QSVD) or the quantum Fourier transform (QFT).

*Quantum Activation Functions*

Quantum activation functions are quantum analogs of classical activation functions in neural networks. They are mathematical functions that transform the output of a quantum circuit into a new quantum state, introducing nonlinearity into the output of a neural network. There are several types of quantum activation functions, including Quantum ReLU (Q-ReLU), Quantum Sigmoid (Q-Sigmoid), and Quantum Softmax (Q-Softmax). Quantum activation functions can be used in quantum machine learning algorithms to improve their performance on specific types of problems.

*Quantum Pooling*

To decrease the spatial dimensions of a quantum feature map while keeping the most important data, the quantum pooling approach is employed in quantum machine learning. Since quantum states are inherently uncertain, it might be challenging to pinpoint the most important details. Numerous quantum pooling strategies, such as quantum maximum pooling, quantum mean pooling and quantum median pooling, have been suggested to overcome this issue. Quantum amplitude pooling is implemented using quantum gates and circuits, such as quantum amplitude pooling. The input quantum state is transformed into the frequency domain using a quantum Fourier transform, and the amplitude of each frequency component is squared using a quantum circuit containing Hadamard gates and Controlled-NOT (CNOT) gates. The squared amplitudes are then used to construct a new quantum state with fewer qubits using a measurement operation.

*Qubit to bit*

Quantum computing involves measuring the state of a qubit to convert it to a classical bit, with the probability determined by the quantum state of the qubit. Quantum convolution is a quantum operation used in quantum neural networks, specifically in Qonvolutional Neural Networks (QCNNs). In Qonvolutional, input data is encoded into a quantum state, usually using qubits and quantum gates are applied to the state to perform the convolution operation. The output is then decoded from the resulting quantum state.

*Task Specific Networks*

For each of the X anchors and Y object classes, the classification subnet forecasts the likelihood of an object's presence at each spatial position. The subnet is a fully connected network interconnected to all levels, sharing features. Its design involves a channel input feature map, four conv layers with filters, ReLU activations, and filters, and sigmoid activations to output binary predictions per spatial location. The object classification subnet and box regression subnet share a common structure, but the box regression subnet uses a class-agnostic bounding box regressor with fewer parameters. This strategy is equally successful, as both subnets employ distinct parameters for regressing the offset from anchor boxes to neighboring ground-truth objects.

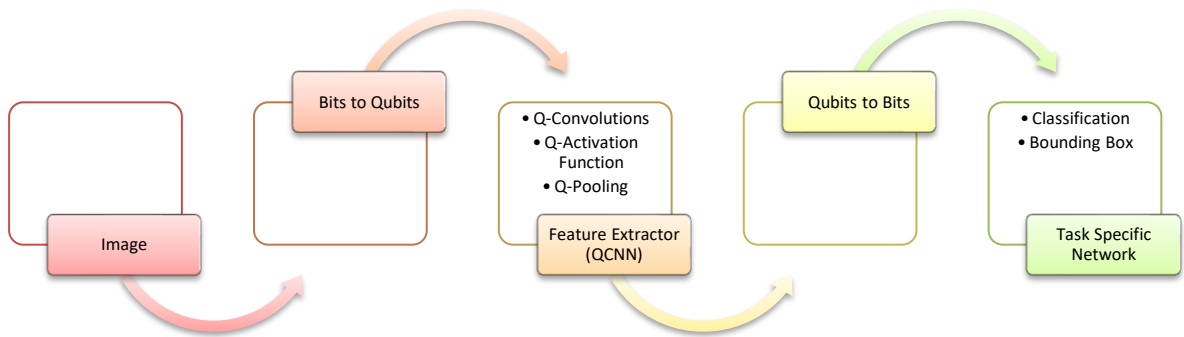

**Figure 3: Process flow diagram of RetinaNet, an image is converted into a vector of corresponding Qubits for quantum computing, feature extraction is performed using QCNN. And Task Specific Networks, draws bounding boxes and detects desired objects.**

Detailed process architecture of the Q-RetinaNet is shown in the fig 3. Here image is fed into the model, then this image is converted to vector of corresponding bits. For quantum computing bits are transformed to Qubits. Once Qubits are fabricated, then feature extraction can be performed using QCNN. QCNN is just like a CNN but only works on Qubits. After extracting the features next step is to forward these feature maps to Task Specific Networks. Due to compatibility issues these Qubits are now need to transformed back to conventional bits. Finally these bits are transferred to Task Specific Networks, which is responsible for drawing bounding boxes and detecting the desired objects from the image.

## Analysis and Results

Various comparative analysis approaches are utilized to compare the performance of the various models employed in the research.

**Accuracy**

The fig 4 illustrates the accuracy of LeNet, AlexNet, VGG, RetinaNet, and Quantum-RetinaNet. Though the Q-RetinaNet contains only fewer Quantum layer but still shows sufficient accuracy, though it is a partial implementation of the QCNN as compared to CNN, however results quite impressive.

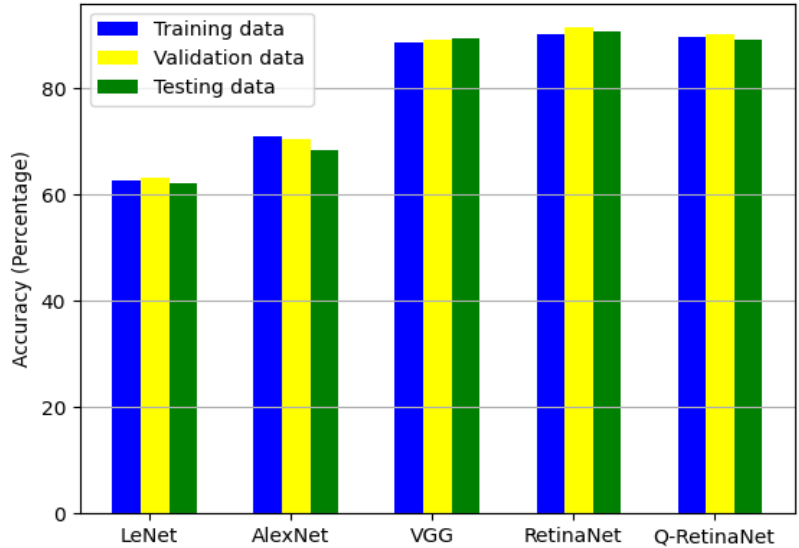

**Figure 4: The comparison of accuracy of LeNet, AlexNet, VGG, RetinaNet, and Quantum-RetinaNet**

**Confusion Matrix**

In machine learning and deep learning, a classification model's performance is assessed using a confusion matrix. It is used to assess the classification model's accuracy by contrasting the anticipated class labels with the actual class labels. An explanation of how the confusion matrix calculates the percentage of true positive (TP), false positive (FP), false negative (FN), and true negative (TN) predictions can be found in the confusion matrix. The instances in each row of the matrix correspond to a predicted class, whereas the examples in each column correspond to an actual class, as shown in fig 5.

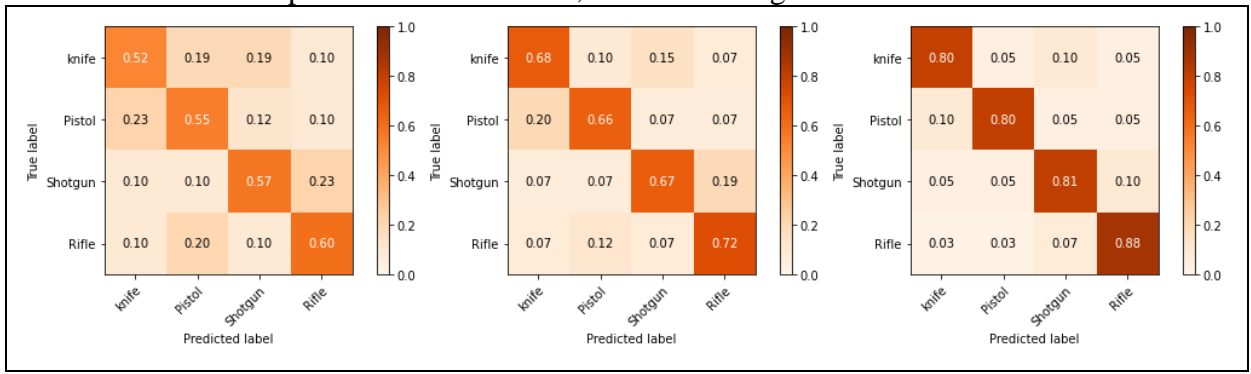

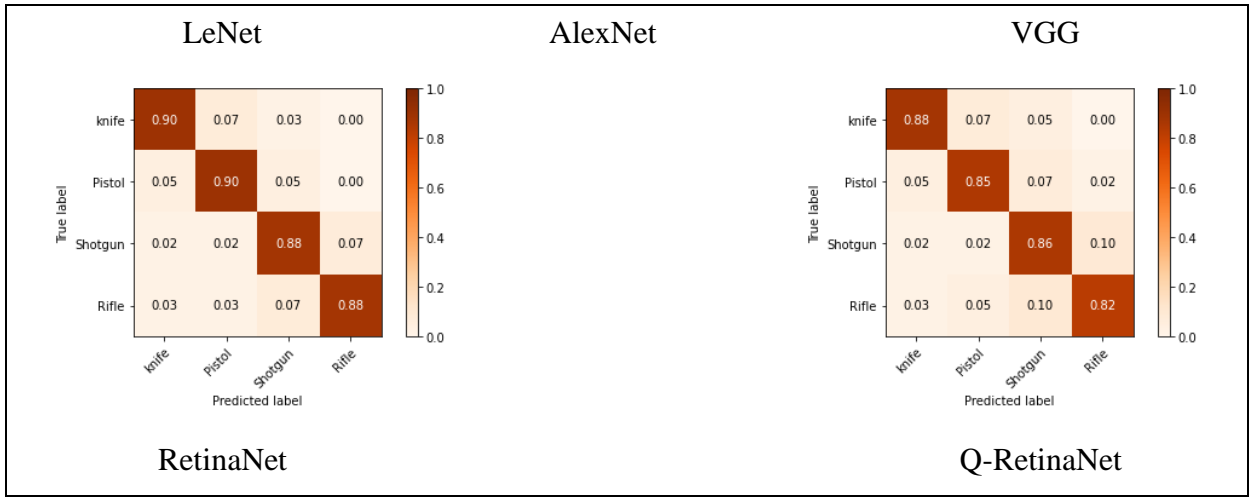

**Figure 5: The comparison of Confusion Matrix of LeNet, AlexNet, VGG, RetinaNet, and Quantum-RetinaNet.**

Several performance indicators are calculated using the confusion matrix, including accuracy, precision, recall, and F1-score. It offers a thorough analysis of the categorization model and aids in determining the model's advantages and disadvantages.

**F1-SCORE**

Deep learning often uses the F1-score to track performance, particularly when it comes to binary classification. It is the harmonic mean of precision and recall, where recall is the ratio of true positives to the total of both true positives and false positives, and precision is the sum of both true positives and false positives (FN). These steps are used to determine the F1 score:

F1-score = 2 * (precision * recall) / (precision + recall)

The F1 score is a crucial indicator for evaluating a classification model's accuracy and recall. It ranges from 0 to 1, with 0 indicating no predictive capacity and 1 indicating perfect precision and recall. It can be computed independently for each class and averaged to provide a performance indicator, see fig 6 for detailed information.

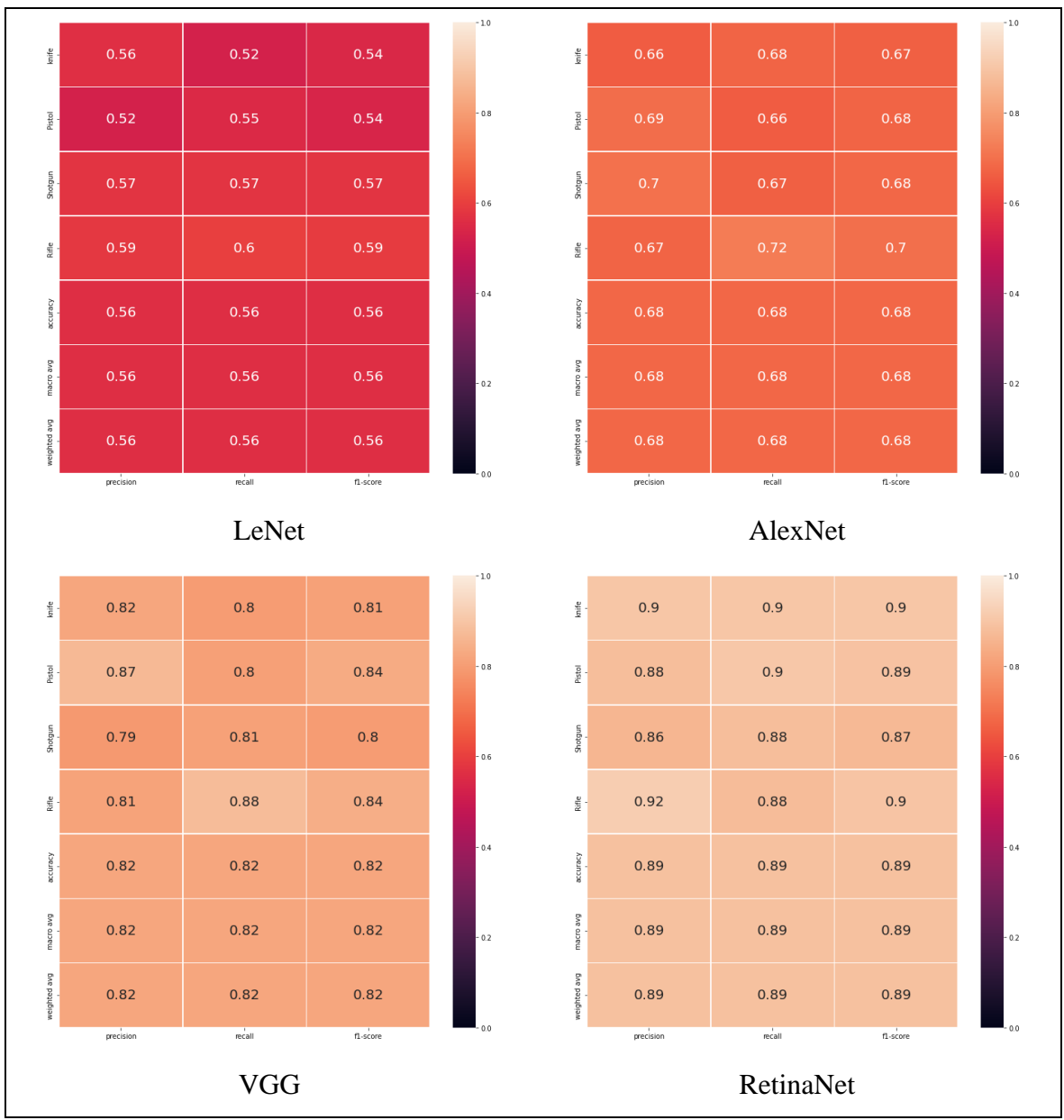

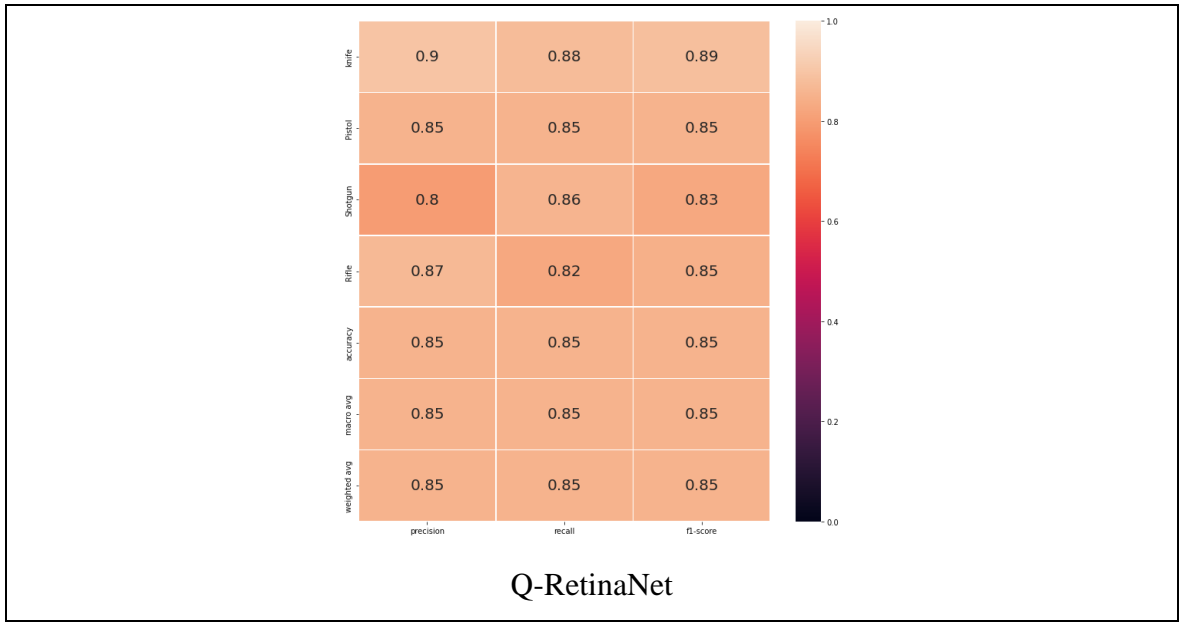

**Figure 6: The comparison of accuracy of LeNet, AlexNet, VGG, RetinaNet, and Quantum-RetinaNet.**

**ROC**

A ROC curve in deep learning measures the accuracy of binary classifier models by plotting the true positive rate (TPR) against the false positive rate (FPR) at different threshold settings, as shown in fig 7. It helps determine the optimal threshold setting and compares the performance of different classifier models. The area under the ROC curve (AUC) is a commonly used metric, with scores ranging from 0 to 1, indicating poor classification and random guessing.

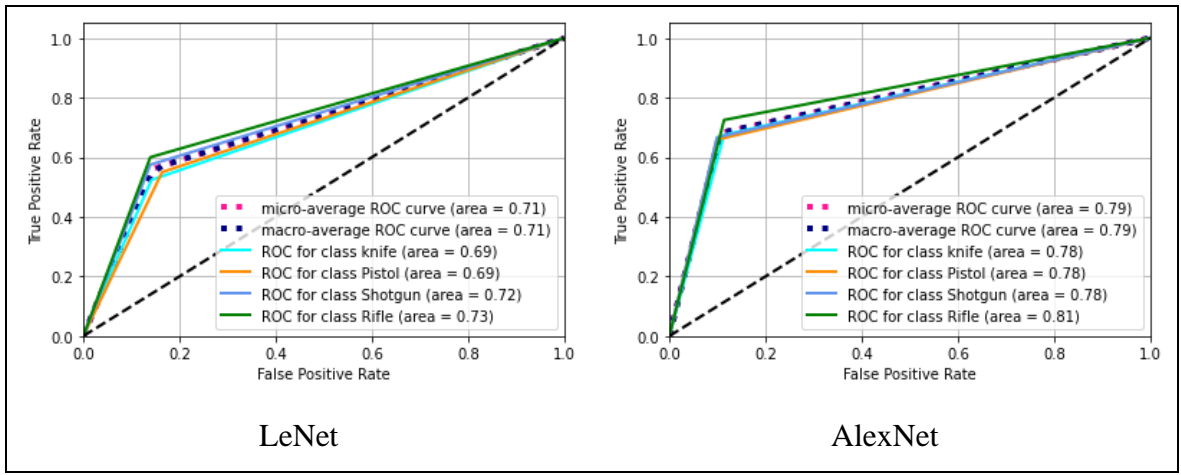

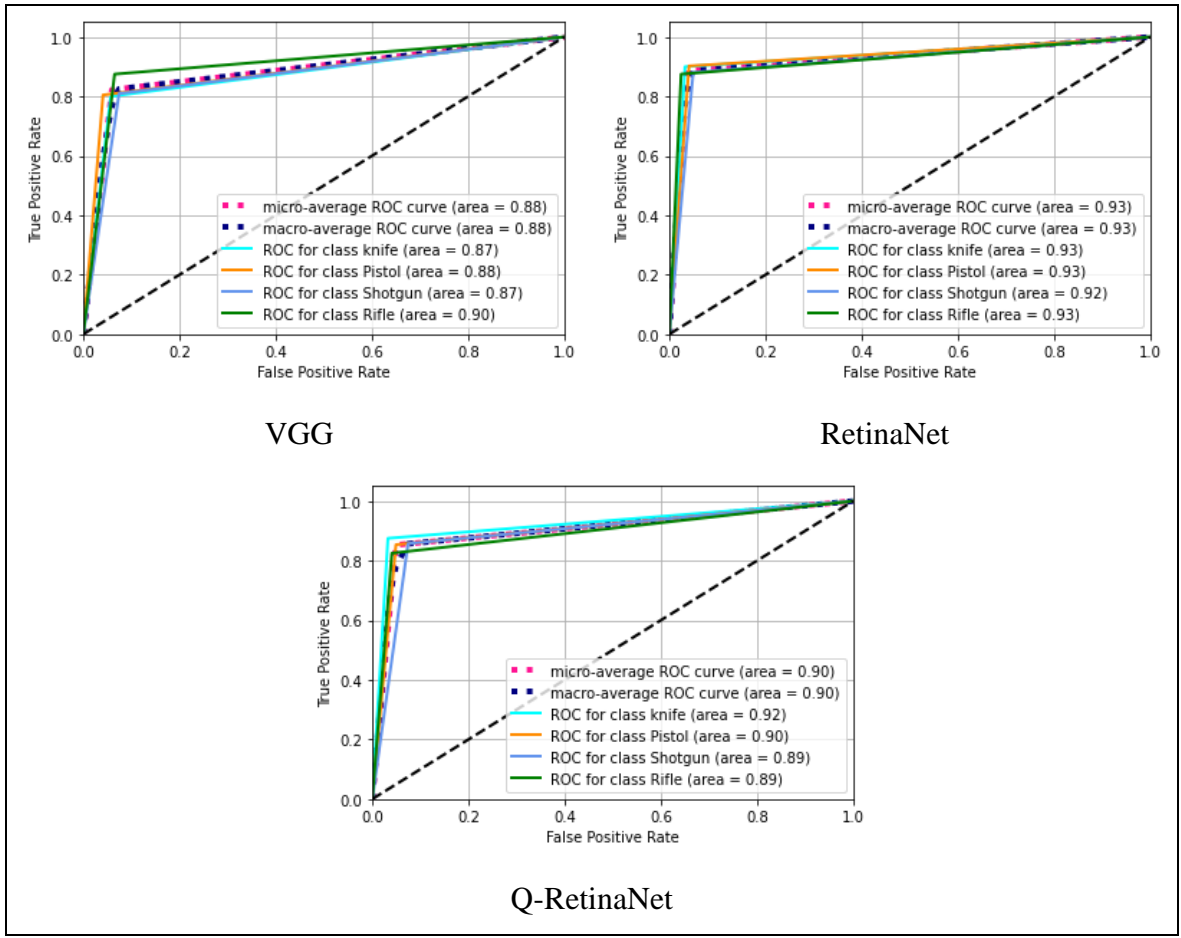

**Figure 7: The comparative analysis of ROC of LeNet, AlexNet, VGG, RetinaNet, and Quantum-RetinaNet.**

## Conclusion

Hybrid neural networks (QCNN-RetinaNet) for object detection integrating conventional and quantum computing methods have attracted a lot of interest. RetinaNet is utilized for object detection, while the QCNN layers are in charge of feature extraction and reduction. Quantum gates and circuits can be used to implement these layers, potentially resulting in exponential speedup in some applications. Traditional deep learning methods like backpropagation and focus loss function may be used to train RetinaNet, which was created to overcome the class imbalance in object recognition. When solving certain object detection issues, the hybrid QCNN-RetinaNet performs better than either a QCNN or RetinaNet by itself. However, in addition to competence in deep learning and neural networks, constructing a hybrid QCNN-RetinaNet combines of the benefits of both classical and quantum computing methods.

## Ethics

A dataset that was generated in this research was designed to have an armory of the types of pistols, shotguns, rifles (limited to short-range), knives, Kalashnikovs, etc. It was defined as a dataset that contained all types of armory most commonly seen in the streets. The dataset shows armories that are commonly found in 3rd world countries, as the project focuses on surveillance in developing countries. This makes the dataset relevant for countries in the Third World. Our investigation included contacting law enforcement, police, private security agencies, websites, the Internet, and the crime branch of a major news channel, among other security departments.

## Conflict of Interest
-The authors declare no competing interests.

## Author contribution statement
Syed Atif Ali Shah (Researcher): Conceptualization idea, methodology, investigation, experiment implementation, data collection, writing. 80%
Dr. Nasir Ageelani (Co-Supervisor): supervision, review. 15%
Dr. Najeeb Al-Sammurrai (Supervisor): supervision and validation. 5%

## Data availability statement
The data supporting the results of this study will be made available by the corresponding author, **Atif**, upon reasonable request.

**quantum deep learning. Indonesian Journal of Electrical Engineering and Computer Science. 2023;30(1):528.**